# Agent-based models of social behaviour and communication in evacuations: A systematic review

**Author names and affiliations**

Anne Templeton[1], Hui Xi[2], Steve Gwynne[2], Aoife Hunt[2], Pete Thompson[2] & Gerta Köster[3]

[1] Department of Psychology, 7 George Square, Edinburgh, EH89JZ, a.templeton@ed.ac.uk
[2] Movement Strategies (a GHD company), 10th Floor, Nexus, 25 Farringdon Rd, London EC4A 4AB
[3] Department of Computer Science and Mathematics, Hochschule Munchen University of Applied Sciences, Munich, Germany

**Highlights:**

- We created four categories of how information is communicated between agents in state-of-the-art agent-based evacuation models.
- We established eight ways that social interactions are implemented in agent-based evacuation models, ranging from low to high social complexity.
- We identified the variables commonly used in agent-based evacuation models and demonstrate their use over recent years.
- We discuss promising avenues forward in how agent-based evacuation models can simulate communication and social behaviour in evacuation based on empirical evidence, and how these might impact commonly used variables in the models.

**Acknowledgements**
We would like to thank Lasse Schaefer, Zhimin Wang, Chamini Gnanavel and Vaibhav Saran for their valuable help during the title and abstract screening of the systematic review.

**Funding source**
This work was supported by the UK Research and Innovation for the project 'Simulating the impact of first responder communication strategies on citizen compliance in emergencies, reference MR/V023748/1.

**Declarations of interest**
None.

Abstract

Most modern agent-based evacuation models involve interactions between evacuees. However, the assumed reasons for interactions and portrayal of them may be overly simple. Research from social psychology suggests that people interact and communicate with one another when evacuating and evacuee response is impacted by the way information is communicated. Thus, we conducted a systematic review of agent-based evacuation models to identify 1) how social interactions and communication approaches between agents are simulated, and 2) what key variables related to evacuation are addressed in these models. We searched Web of Science and ScienceDirect to identify articles that simulated information exchange between agents during evacuations, and social behaviour during evacuations. From the final 70 included articles, we categorised eight types of social interaction that increased in social complexity from collision avoidance to social influence based on strength of social connections with other agents. In the 17 models which simulated communication, we categorised four ways that agents communicate information: spatially through information trails or radii around agents, via social networks and via external communication. Finally, the variables either manipulated or measured in the models were categorised into the following groups: environmental condition, personal attributes of the agents, procedure, and source of information. We discuss promising directions for agent-based evacuation models to capture the effects of communication and group dynamics on evacuee behaviour. Moreover, we demonstrate how communication and group dynamics may impact the variables commonly used in agent-based evacuation models.

## 1    Introduction

Agent-based models of evacuation behaviour are used to simulate and plan how people egress from numerous environments in emergencies. They can be used to simulate previous evacuations, predict how people will evacuate in various scenarios, compare emergency procedures, and advise on the practicality and safety of evacuation plans for real-world incidents. However, a challenge in developing agent-based evacuation models is to ensure their assumptions for decision-making and behaviour are valid and that their simulations provide as realistic an approximation as possible to actual behaviour. Many agent-based evacuation models incorporate factors that research has shown guide decision-making in evacuations, such as perceived risk to the self (Kinateder et al., 2015), lighting conditions (Cosma et al., 2016), and physical ability (Geoerg et al., 2019). These factors are certainly important, but they tend to focus on perception and actions at the individual level. Evacuations frequently involve multiple people, such as colleagues, family members, or fellow attendees at an event. As such, it is important to develop an understanding of the influences between agent interactions and emergent, collective group dynamics and behaviours.

### 1.1. Group processes and communication in evacuations

Emergency evacuations typically require more than one person to evacuate, particularly in public spaces, multi-household buildings, and at organised events. Previous research suggests that responses of people to emergencies is diverse, but group processes consistently influence how people evaluate situations and decide to act. For example, people tend to attempt to find members of their group when they are separated (Sime, 1985). In ambiguous situations, people look to others as a diagnostic tool to decide how to act (Drury et al., 2023). In particular, people seek information from those with whom they are already familiar or perceive to be in the same group as them. If the people experiencing the emergency are strangers, they psychologically come together as a group and help one other on the basis of being part of the same group (Drury et al., 2009a). This can lead to large-scale coordination such as helping injured people to evacuate, as well as providing emotional and practical support to others (Drury et al., 2015; Ntontis et al., 2018).

Other important group processes during evacuations are how the relationships between the information providers and receivers influence their willingness to follow evacuation recommendations. Research from social psychology suggests that people are more highly influenced by others in their group because they believe them to be acting in their shared best interests (Reicher et al., 2010). For example, residents of high-rise buildings were more motivated to engage with evacuation guidance for their building if they believed it came from a person or organisation who was acting in the best interests of the residents, such as fire and rescue services (Templeton et al., 2023a).

Although prior relations between responders and the public are important, the information that responders provide the public about incidents can influence the public's view of them during an emergency. This is important because the changing views of the information providers may dynamically influence behaviour throughout the evacuation. For example, Carter et al. (2015) found that the public felt responders were more legitimate and had a stronger sense of being in the same group with them when they provided information about the emergency, explained the actions required and why they were important, and gave regular updates. In response, people were more willing to comply with the instructions for undergoing a decontamination process.

Taken together, the research from social psychology suggests that people in evacuation scenarios share information with each other and react to information differently depending on their views of the provider of the information. Thus, we argue that the social relations and communication between agents are important factors to understand how behaviour evolves within agent-based evacuation models.

### 1.2. Synthesising variables used in agent-based evacuation models

The purpose and focus of agent-based evacuation models vary. For example, models may aim to re-enact simulations of behaviour in previous evacuations (e.g., D'Orazio et al., 2014; Makinoshima et al., 2022; Takabatake et al., 2020), predict the effects of environmental conditions such as the effect of door width on evacuation time (e.g., Şahin et al., 2019) or obstacles on exit choice (Zhou et al., 2021), or simulate the effect of personality types on evacuation behaviour (e.g., Li & Han, 2015; Zhang et al., 2017). The type of statistical results measured following simulation will depend on the model purpose. For example, a model measuring escape efficiency may measure density levels at specific locations (e.g., Wang & Jiang, 2019). A model which aims to simulate how people follow a leader's behaviour may measure how long it takes agents to evacuate (e.g., Li & Han, 2015), or the number of exited agents who reach an exit (e.g., Fang et al., 2016).

Despite the varied aims and purposes of the models, all manipulate and measure the effects of variables related to evacuation behaviour. Establishing the key variables manipulated and measured in agent-based evacuation models can show key themes and areas of interest in the literature. However, it can also help to identify which prominent variables the social psychology literature suggests are influenced by group processes and information sharing. For example, how likely agents are to respond to information from others depending on their views of the information givers, or how seeking information or other people may impact pre-evacuation delay.

### 1.3. Current study

Research on social behaviour in emergencies has been evolving for decades, including influential papers such as Johnston and Johnson's (1989) research the 1977 fire at Beverly Hills Support Club, and Johnson's (1987) investigation into the 1979 The Who concert disaster. Both papers show the prominence of helping behaviour during the disasters. Similarly, research on disaster communication is well established, since prominent papers such as Reynolds and Seeger (2007) have posed effective models of crisis communication that could be applicable to evacuations. However, we argue that it is now necessary to systematically review the extent to which research in these areas are included in agent-based evacuation models. The systematic pattern of findings in social psychology specifically about group processes in evacuations started to gain prominence in 2009 and have continually developed since. More recently, the important inter-relations between communication approaches and group processes in emergencies has gained prominence. As the field has rapidly developed, so too have efforts to develop interdisciplinary collaboration with evacuation modellers (for an example of these efforts, see Adrian et al., 2019; Gwynne & Hunt, 2018; Scholz et al., 2023).

The extent to which the more recent findings on group processes on communication are reflected in agent-based evacuation models is unclear. As such, we aimed to systematically review the publicly available literature of existing agent-based evacuation models, focusing on how the models conceptualise social interactions and communication between agents before and during evacuations. In doing this, we aimed to identify the current abilities of models to simulate social behaviour in evacuations, particularly the interactions between first responders and the public during evacuations. Following this, we aimed to synthesise the variables used in agent-based evacuation models and identify how these may be related to our primary focus on group dynamics and communication.

In addition to this review, we conducted an informal review of commercial software that simulates evacuation behaviour to determine their capabilities to model social interactions and communication. These capabilities would not necessarily be captured in the primarily academic literature, so this was a supplementary objective of the review and required a more ad hoc approach than the primary systematic review. The supplementary review is attached to this article in Appendix A.

Research questions for the primary review were:

1) How do current agent-based models simulate social interactions and communication approaches between first responders and the public in evacuations?
2) What are the common variables addressed in these models?

## 2    Methodology

The planned methodology for this systematic review was pre-registered on the Open Science Framework (https://osf.io/hqk54). The full dataset of included articles is also available at https://osf.io/96k7q/files/osfstorage/651fe4eb0ee23200be41a01f.

### 2.1. Design

The systematic review involved four stages. First, we searched and identified relevant agent-based evacuation models that modelled social behaviour and communication between agents. Second, we characterised the literature by creating categories according to how social interactions between agents were implemented. Third, we created categories according to how communication between agents was implemented. Finally, we identified the main independent and dependent variables used in the models and synthesised these into categories according to their area of focus and tracked their prevalence over time. The details for the supplementary commercial software review are included in Appendix A.

### 2.2. Databases and inclusion criteria

We searched the electronic databases *Web of Science* and *ScienceDirect* for articles that presented the capabilities of their models to simulate social behaviour and communication in evacuations. The inclusion and exclusion criteria are presented in Table 1.

Table 1: Inclusion and exclusion criteria used in the systematic review.

| Inclusion criteria | |
|---|---|
| 1. | The simulation of a public response to emergency evacuation scenarios, including the need for people to defend in place or relocate. |
| 2. | The simulation of information exchange between agents during evacuations (particularly verbal communication). |
| 3. | The simulation of social behaviour during evacuations (e.g., affiliative, grouping, helping or other coordination behaviour) reflecting role-based effects (e.g., first responders) where appropriate. |
| 4. | How social behaviour and/or communication are simulated. This may include but not be limited to:<br>a) Agents can store and share information;<br>b) Agents process current/stored information as part of their decision-making;<br>c) Agents can be affected by actions of other agents;<br>d) Agents can have different roles;<br>e) Roles can affect actions;<br>f) Roles can affect group membership;<br>g) Group membership can affect information sharing;<br>h) Group membership can affect actions. |
| **Exclusion criteria** | |
| 1. | Not publicly available agent-based models. |
| 2. | Not in English. |
| 3. | Provided no description of the underlying conceptual model. |
| 4. | Published from 2010 to 2022* to ensure we capture current trends. |

\* The search was conducted in August 2022, thus no papers published later than that are included.

### 2.3. Search terms and strategy

The search strategy was developed for *Web of Science* and then adapted for *ScienceDirect* to fit its search string requirements which allow no more than eight Boolean operators. We identified the terms frequently used to describe social behaviour and communication between occupants during evacuations. Then, we constructed the search strings into three stages to refine the relevance of the search (see Figure 1):

    1) Established a focus on agent-based modelling (e.g., using the term 'agent'),

    2) Established scenarios (e.g., using the term 'emergency'), and

    3) Established modelling capabilities (e.g., e.g., 'communication, 'first responder').

The search strings for both *Web of Science* and *ScienceDirect* across the three stages are shown in Table 2.

Figure 1: Search string construction.

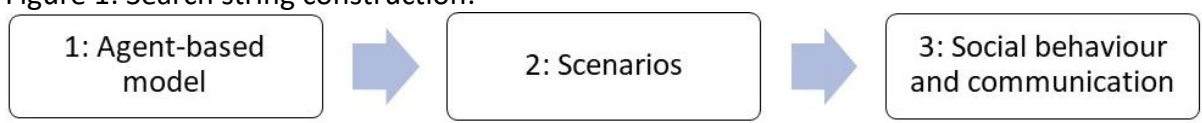

Table 2: Search strings used in *Web of Science* and *ScienceDirect*.

| | **Purpose** | *Web of Science / Science Direct* ['Search terms'] | |
|---|---|---|---|
| 1. | **Agent-based modelling:** Specifies the search range to be within agent-based models. | | |
| | *Web of Science* | '(model OR simulat* OR micro OR agent OR occupant)' | |
| | *ScienceDirect* | '(model/modelling OR simulation) AND (agent OR occupant OR evacuee)' | |
| 2. | **Scenarios:** Specifies two modelling scenarios: crowd movement and emergency evacuation in which communication take place. | | |
| | *Web of Science* | '(crowd AND dynamic OR movement AND emergency OR evacuat' | |
| | *ScienceDirect* | '(crowd AND (dynamic OR movement)) OR emergency OR evacuation' | |
| 3. | **Social behaviour and communication:** | | |
| | Group terms | Seeks models that represent group and/or social behaviour among agents. | |
| | | *Web of Science* | 'Social AND behaviour' or 'Group AND behaviour' |
| | | *ScienceDirect* | 'social OR group AND behaviour' |
| | Communication | Seeks models that include first responder interventions: communication, interaction or intervention. | |
| | | *Web of Science* | N/A - scoping produced irrelevant results. |
| | | *ScienceDirect* | 'communication OR interaction OR intervention' |
| | Roles | Seeks models that represent agents with different roles and the capability to carry out various tasks. | |
| | | *Web of Science* | 'role', 'action' |
| | | *ScienceDirect* | 'role OR task OR action' |
| | First responder roles | Search for any explicit representation of the specific roles (e.g., first responder, fire and rescue services) and their interaction with the other agents. | |
| | | *Web of Science* | 'emergency management' |
| | | *ScienceDirect* | '"first responder" OR "fire fighter" OR "emergency management" OR rescue OR police OR ambulance' |
| | Instructions and guidance | Seeks models that represent agent adherence to the instructions received from first responders. | |
| | | *Web of Science* | 'guidance' and 'order' |
| | | *ScienceDirect* | 'instruction OR guidance OR compliance OR order' |

## 2.4. Filters and limits

We refined the search results by filtering according to broad subject areas (see Table 3). In *ScienceDirect*, we further filtered by the most relevant common journals. This list of journals is provided in Appendix B.

Table 3: Search results on 30th August 2022 prior to title and abstract screening.

| Platform | Subject Areas | Results prior to screening |
|---|---|---|
| *Web of Science* | Physics (e.g., multidisciplinary), Engineering (e.g., civil, multidisciplinary, software) Computer Science (e.g., computer science interdisciplinary applications, computer science artificial intelligence, computer science information systems), Transportation science technology, Environmental studies, Mathematics (e.g., multidisciplinary), Psychology (e.g., applied), Multidisciplinary Sciences, Public environmental occupational health, Social sciences interdisciplinary | 28 (after 20 duplicates removed) |
| *Science Direct* | Computer Science, Engineering, Mathematics, Decision Sciences, Social Sciences, Psychology, (Earth and Planetary Sciences, Physics and Astronomy, Environmental Science, Neuroscience) | 661 (after 467 duplicates removed) |
| | | Total = 689 |

## 2.5. Data extraction

We used Covidence (https://www.covidence.org/), a tool to assist systematic reviews, to collate the search results and remove duplicates. For the initial title and abstract screening, we used Excel to track and code the literature, noting where articles were excluded and the reason. During the full screening stage, we imported the author(s), year of publication, and DOIs into our spreadsheet then manually recorded the following information: model name or models used (if relevant), type of model (e.g., decision-making evacuation model, pedestrian dynamics model), core variables included as independent and dependent variables, category of social interactions between agents, and the category of communication between agents. The final stage of our data synthesis was to categorise in the spreadsheet the commonly used variables in the agent-based evacuation models, according to whether they were independent variables (i.e., something manipulated in the model) or dependent variables (i.e., outcome variables in reaction to what was manipulated).

## 3   Results

A total of 70 articles were included in the full text review after applying our inclusion and exclusion criteria. We then created categories of the nature of the agents' social interactions and how communication between agents was implemented, and then synthesised the most commonly used variables in the models (see Figure 2).  The full list of included articles and their respective categories are shown in Table 4. In the supplementary materials, we also provide a full list of details for the articles, including the author(s), year of publication, title, relevant model names, theoretical basis for the models, and categories of social interactions, communication, and common variables used. The results for the supplementary review of the commercial software are included in Appendix A.

Figure 2: Systematic review process.

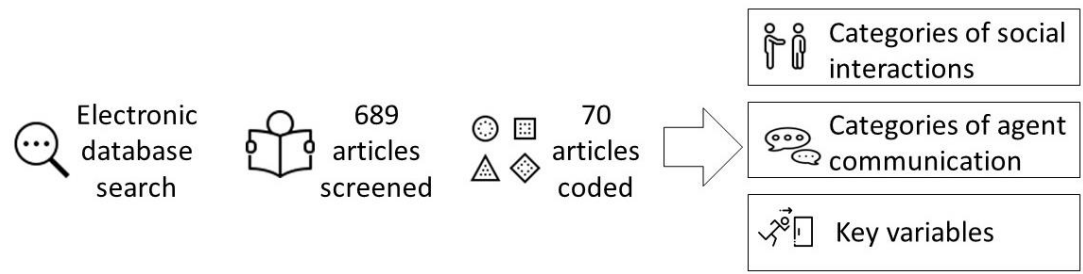

Table 4: List of articles included in the review.

| Author(s) | Year | Social interaction category | Communication category | Variable categories |
|---|---|---|---|---|
| Aurell & Djehiche | 2019 | Maintaining the group structure | N/A | Personal attributes; procedure; Environmental condition |
| Bao & Huo | 2021 | Collision avoidance | N/A | Personal attribute; Environmental condition |
| Barnes et al. | 2021 | Maintaining the group structure | N/A | Personal attribute |
| Bernardini et al. | 2014 | Maintaining the group structure | Spatial location: area around agent | Environmental condition; Procedure; Information; Personal attribute |
| Best et al. | 2014 | Collision avoidance | N/A | Personal attributes; Procedure |
| Bulumulla et al. | 2022 | Influence via information transfer | Spatial location: area around agent; Social network links | Personal attributes; Information; Procedure |
| Cao et al. | 2017 | Influence via emotion transfer; Following behaviour | N/A | Personal attribute; Procedure |
| Cao et al. | 2021 | Competition between agents; Influence via visual perception | N/A | Personal attributes; Procedure; Environmental condition |
| Chen et al. | 2021 | Influence via emotion transfer; Following behaviour | N/A | Personal attributes; Environmental condition; Procedure |
| Cheng & Zheng | 2018 | Competition between agents | N/A | Personal attributes; Environmental condition; Procedure |

| Author(s) | Year | Social interaction category | Communication category | Variable categories |
|---|---|---|---|---|
| Delcea & Cotfas | 2019 | Collision avoidance | N/A | Personal attributes; Procedure; Environmental condition |
| Ding | 2011 | Following behaviour | N/A | Personal attributes; Procedure |
| D'Orazio et al. | 2014 | Maintaining the group structure | N/A | Personal attributes; Procedure; Information; procedure |
| D'Orazio et al. | 2014 | Maintaining the group structure | N/A | Environmental condition; Personal attributes; Procedure |
| Dossetti et al. | 2017 | Collision avoidance | N/A | Personal attributes; Procedure; Environmental condition |
| Fang et al. | 2016 | Following behaviour; Competition between agents | N/A | Personal attributes; Environmental condition |
| Fu et al. | 2014 | Influence via emotion transfer | N/A | Procedure; Environmental condition |
| Fu et al. | 2013 | Collision avoidance | N/A | Personal attributes; Procedure; Environmental condition |
| Gao et al. | 2022 | Following behaviour | N/A | Procedure; Environmental condition |
| Gao et al. | 2020 | Collision avoidance; Following behaviour | N/A | Personal attributes; Procedure |
| Guan et al. | 2016 | Competition between agents | N/A | Personal attributes; Procedure; Environmental condition |
| Haghani & Sarvi | 2019 | Influence via visual perception | N/A | Personal attributes; Procedure; Environmental condition |
| Harris et al. | 2022 | N/A | External communication | Personal attributes; Procedure |
| Hasan & Ukkusuri | 2011 | Influence via pre-existing bonds | Social network links | Personal attributes; Procedure |

| Author(s) | Year | Social interaction category | Communication category | Variable categories |
|---|---|---|---|---|
| Heliövaara et al. | 2012 | Collision avoidance | N/A | Environmental condition; Procedure |
| Henein & White | 2010 | Influence via information transfer; Influence via visual perception | Spatial location: area around agent | Personal attributes; Procedure; Environmental condition |
| Huang et al. | 2022 | Influence via information transfer; Influence via emotion transfer. | N/A | Personal attributes; Procedure; Environmental condition; Information |
| Jiang et al. | 2014 | Influence via visual perception; Following behaviour | N/A | Personal attributes; Procedure |
| Kasereka et al. | 2018 | Influence via visual perception | N/A | Procedure; Environmental condition; Information |
| Kim et al. | 2018 | Collision avoidance | N/A | Personal attributes; Environmental condition |
| Lei et al. | 2012 | Following behaviour; Collision avoidance | N/A | Personal attributes; Procedure; Environmental condition |
| Li & Han | 2015 | Collision avoidance; Influence via information transfer | N/A | Personal attributes; Environmental condition; Procedure |
| Li & Qin | 2012 | Influence via emotion transfer; Following behaviour | Spatial location: information trails | Personal attributes; Procedure; Environmental condition |
| Li et al. | 2019 | Influence via information transfer | Spatial location: area around agent | Personal attributes; Procedure; Environmental condition; Information |
| Lopez-Carmona & Garcia | 2022 | Following behaviour | N/A | Personal attributes; Procedure; Environmental condition |
| Lopez-Carmona & Garcia | 2021 | Collision avoidance | N/A | Environmental condition; Procedure |

| Author(s) | Year | Social interaction category | Communication category | Variable categories |
|---|---|---|---|---|
| Lovreglio et al. | 2016 | Influence of information transfer; Influence via pre-existing bonds | Spatial location: area around agent | Personal attributes; Information; |
| Müller et al. | 2014 | Following behaviour; Maintaining the group structure | N/A | Personal attributes; Procedure; Environmental condition |
| Makinoshima et al. | 2022 | Influence via information transfer | Spatial location: area around agent; Social network links | Personal attributes |
| Marzouk & Daour | 2018 | Collision avoidance | N/A | Procedure |
| Mesmer & Bloebaum | 2014 | Competition between agents | N/A | Personal attributes; Procedure |
| Mohd Ibrahim et al. | 2017 | Competition between agents | N/A | Personal attributes; Procedure; Environmental condition |
| Niu et al. | 2021 | Competition between agents; Influence via emotion transfer; Influence via visual perception | N/A | Personal attributes; Procedure; Environmental condition |
| Poudel et al. | 2018 | Collision avoidance | N/A | Personal attributes; Procedure |
| Ramírez, et al. | 2019 | Collision avoidance | N/A | Personal attributes; Procedure; Environmental condition |
| Rigos et al. | 2019 | Influence via information transfer | External information provided | Personal attributes |
| Şahin et al. | 2019 | Following behaviour | N/A | Personal attributes; Environmental conditions |
| Serrano & Botia | 2013 | Influence via visual perception | N/A | Personal attributes; Procedure; Environmental condition; Information |

| Author(s) | Year | Social interaction category | Communication category | Variable categories |
|---|---|---|---|---|
| Song et al. | 2019 | Collision avoidance; Influence via visual perception | N/A | Personal attributes; Procedure; Environmental condition |
| Srinivasan et al. | 2017 | Following behaviour; Influence via visual perception | Spatial location: area around agent | Personal attributes; procedure; Environmental condition |
| Takabatake et al. | 2020 | Maintaining the group structure | Spatial location: area around agent | Personal attributes; Procedure |
| Takabatake et al. | 2017 | Influence via visual perception | N/A | Personal attributes; procedure; Environmental condition |
| Tan et al. | 2015 | Influence via visual perception | N/A | Personal attributes; Environmental condition; Procedure |
| Tang et al. | 2015 | Collision avoidance | N/A | Personal attributes; Procedure; Environmental condition |
| Tinaburri | 2022 | Influence based on pre-existing bonds | Social network links | Personal attributes; Procedure |
| Tissera et al. | 2013 | Collision avoidance | N/A | Personal attributes; Procedure |
| Tissera et al. | 2012 | Collision avoidance | N/A | Personal attributes; Procedure; Environmental condition |
| von Schantz & Ehtamo | 2022 | Influence via information transfer | N/A | Procedure; Environmental condition |
| Wang et al. | 2015 | Following behaviour; Influence via visual perception | N/A | Personal attributes; Procedure; Environmental condition |
| Wang & Jiang | 2019 | Collision avoidance | N/A | Personal attributes; Procedure |
| Wang et al. | 2012 | Influence via information transfer | Spatial location: information trails | Personal attributes; Procedure |
| Yang et al. | 2016 | Following behaviour | N/A | Personal attributes; Procedure |

| Author(s) | Year | Social interaction category | Communication category | Variable categories |
|---|---|---|---|---|
| Yang et al. | 2019 | Influence via pre-existing bonds | Social network links | Personal attributes; Procedure; Information |
| Yue et al. | 2011 | Influence via visual perception | N/A | Personal attributes; Procedure; Environmental condition |
| Zhang et al. | 2014 | Influence via information transfer | Spatial location: information trails | Personal attributes; Environmental cues; Risk awareness; Source of information |
| Zhang et al. | 2022 | Competition between agents | N/A | Environmental condition; Personal attributes; Procedure |
| Zhang et al. | 2022 | Influence via visual perception | N/A | Personal attributes; Information; Procedure |
| Zheng et al. | 2019 | Following behaviour; Influence via visual perception. | Spatial location: information trails | Personal attributes; Procedure; Environmental condition |
| Zhou et al. | 2021 | Influence via visual perception; Collision avoidance | N/A | Personal attributes; Environmental condition |
| Zlateski et al. | 2020 | Influence via visual perception | N/A | Environmental condition; Procedure |

### 4.1. Synthesis of social interactions in agent-based evacuation models

We identified eight categories of how social interactions between agents were implemented. These ranged from calculating collision avoidance to including complex social ties and influence. The first four categories tend to model physical reasons for social behaviour (e.g., solely avoiding collision, focusing on maintaining group structure), and the latter four categories include more complex social influence on decision-making based on the actions of other agents and social connections with them (see Figure 3).

Notably, 18 models included two forms of social behaviour, one model included three forms of social behaviour (Yunyun et al., 2012), and one article (Harris et al., 2022) did not model social behaviour but was included in the overall review due to their formalisation of communication between agents.

Figure 3: Overview of social interaction categories in agent-based evacuation models.

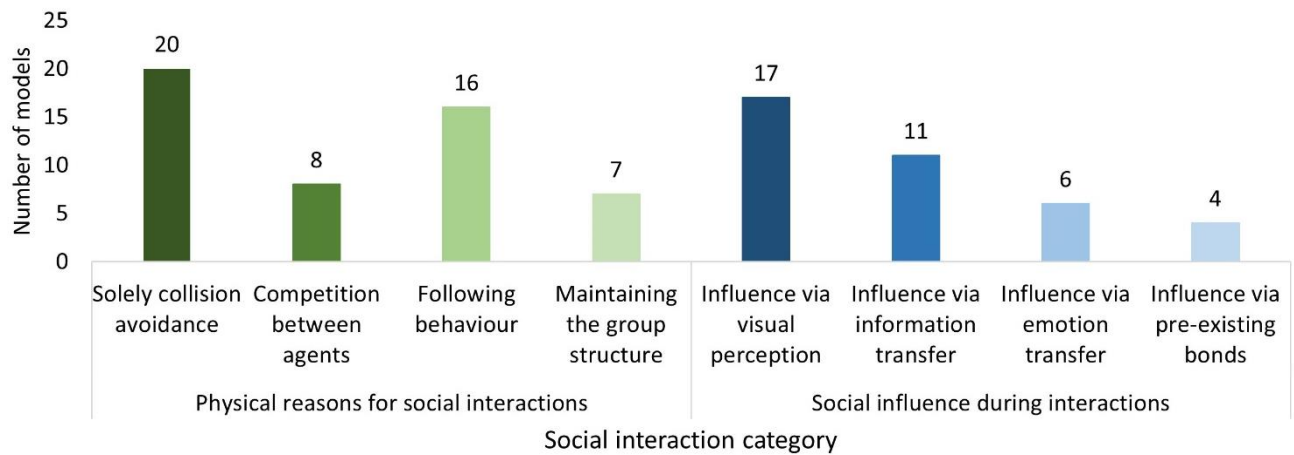

### 4.1.1. Solely collision avoidance

Articles in this category (*n* = 20, 28.5% of total articles modelling social interactions) solely focused on agents avoiding colliding or overlapping with other agents in the environment and did not include any other social interactions. For example, agents were guided by repulsion forces (e.g., Heliovaara et al., 2012), or random choice to decide which agent moved into a space (e.g., Tissera et al., 2013), or give priority to agents with faster movement speed (e.g., Tissera et al., 2012).

### 4.1.2. Competition between agents

In these models (*n* = 8, 11.4%), agents explicitly competed for space in the same area or cell, and the model primarily focused on the results or resolution of the conflict. Ways of resolving conflict included using game theory (e.g., Guan et al., 2016) and having agents with different personality types such as cooperative or competitive wherein the competitive agents would get priority (e.g., Cheng & Zheng, 2018).

### 4.1.3. Following behaviour

Typically categorised by leader-follower models, these articles (*n* = 16, 22.9%) had follower agents be attracted to the location of leader agents and then tended to stay with them throughout the simulation (e.g., Lopez-Carmona & Garcia, 2022), or would follow other agents who were in front of them (typically referred to in the models as 'herding' behaviour) (e.g., Ding, 2011).

### 4.1.4. Maintaining group structure

In these models (*n* = 7, 10%), agents were put into groups with others and attempted to move together with their group throughout the simulation. For example, agents gravitated towards the centre of their group's structure as they navigated the evacuation (e.g., D'Orazio et al., 2014), or decreased the speed of the group to match the slowest member as they evacuated (e.g., Takabatake et al., 2020).

### 4.1.5. Influence via visual perception

Models in this category ($n$ = 17, 24.3%) focused on how agents' decision-making during the evacuation was influenced by the behaviour of other agents within their visual field. For example, agents may have perceived others leaving the environment and then decided to evacuate themselves (e.g., Kasereka et al., 2018).

### 4.1.6. Influence via information transfer

Articles in this category ($n$ = 11, 15.7%) modelled how agents shared information with others and this guided their evacuation decisions. For example, agents may have informed others about risks in the environment (e.g., Zhang et al., 2014), or knowledge of the environment including evacuation routes (e.g., Li & Han, 2015) which then influenced their behaviour during the evacuation. Further information about how this information was transferred between agents is described in section 4.2.

### 4.1.7. Influence via emotion transfer

Categorised by emotion transference between agents, these models ($n$ = 6, 8.6%) instantiated the ability for agents to influence the emotional state of other nearby agents. These models predominantly focused on 'panic' transfer between agents and how it affected their evacuation. For example, a set proportion of agents were allocated a 'panicked' state at the start of an emergency and then caused other agents to become panicked when in close proximity to them (e.g., Chen et al., 2021). The panicked agents then attempted to speed up their evacuation (e.g., Cao et al., 2017), therefore effecting their evacuation behaviour.

### 4.1.8. Influence via pre-existing bonds

These models ($n$ = 4, 5.7%) were characterised by agents' choices and behaviour being influenced by other agents in their group. For example, household groups were linked by a social network to other households (not necessarily physically nearby) and decided to evacuate once a certain percentage of households in their network had decided to evacuate (Yang et al., 2019). Importantly, agents in some models were most influenced by the evacuation decisions of those with whom they had the strongest relationships. For example, agents had set weights (strength) of connections with other agents and were most influenced to evacuate when those with the strongest weights evacuated (Hasan & Ukkusuri, 2011).

### 4.1.9. Prevalence of social categories over time

The number of models published between 2010 and 2022 in each category of social interaction is shown in Figure 4. There are a low number of publications within each category per year, but there does not appear to be any significant changes in popular approaches over recent years in the data that is available. Possible notable exceptions are the 'influence via information transfer approach' category, which was most widely used in in 2022, and the 'influence via visual perception' and 'collision avoidance' categories were most popular in 2019 and 2021. A table of the number of models within each category across the years is provided in Appendix C.

Figure 4. Timeline depicting the number of models using each category of social interaction from 2010 to 2022.

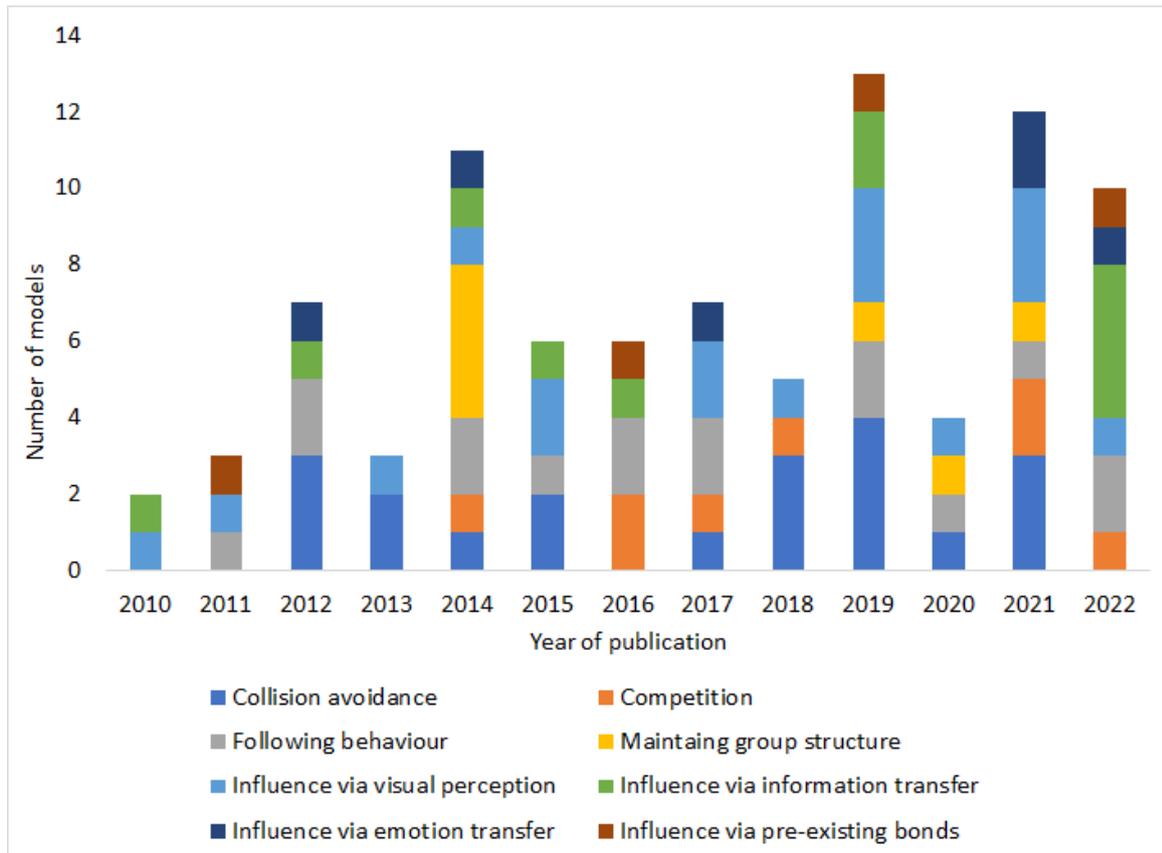

## 4.2. Synthesis of communication between agents in agent-based evacuation models

We categorised the 17 articles which modelled communication between agents based on how the communication was implemented (see Table 5). According to the model descriptions, communication was implemented either based on physical location, through networks, or externally to agents. Notably, the communication between agents in Makinoshima et al. (2022) and Bulumulla et al. (2022) fell into two categories since communication was implemented in multiple ways.

### 4.2.1. Spatial location: Area around agent

Models in this category (*n* = 8) were characterised by 'informed' agents having a field of influence around them (such as a radius) which communicated information to other agents within that radius. In some models, the strength of the communicated information to others decreased as the distance from the agent within the area increased (e.g., Srinvasan et al., 2017). When agents decided whether to evacuate, information communicated from certain agents may have been more influential, such as because they were credible sources of information (e.g., Li et al., 2019), or members of the same group (e.g., Lovreglio et al., 2016).

Table 5: Categories of communication.

| Category | Implementation of communication | No. of articles | Percentage |
|---|---|---|---|
| Spatial location: area around agent | Agents have a bounded field of influence around them which transmits information to other agents. | 8 | 47.1% |
| Social network links | Agents share information to others within their social network. | 5 | 29.4% |
| Spatial location: information trails | Agents disseminate information via trails in the environment as they move. | 4 | 23.5% |
| External communication | Agents in particular areas become informed by an external source as the simulation develops. | 2 | 11.8% |

### 4.2.2. Social network links

This category included five articles where the models were characterised by agents communicating information via social networks. For example, agents in social networks communicated whether they were staying or evacuating (Yang et al., 2019). In some models, agents could be part of multiple networks (e.g., Hasan & Ukkusuri, 2011) and have stronger relationships with certain agents in their networks than others (e.g., Bulumulla et al., 2022). The decisions to evacuate could be dependent on how many others in the network communicated that they had decided to evacuate (e.g., Yang et al., 2019).

### 4.2.3. Spatial location: Information trails

In these models ($n = 4$), 'informed' agents would leave trails of information about either the environment or the source of danger as they moved through the environment, and other agents who moved into the trail would become informed (e.g., Zhang et al., 2014). In some models, agents became increasingly informed the more they encountered trails of information (e.g., Wang et al., 2012). Whether or not the agents were informed typically affected their evacuation behaviour, such as in Li and Qin (2012) where agents informed of the risk become 'panicked' and increased their evacuation speed.

### 4.2.4. External communication

Two models included communication of information about the emergency to agents from an external source. In Harris et al. (2022), evacuation orders were communicated by emergency management to agents according to their locational on the floor field, i.e., agents in particular locations became informed about the emergency from an external source. The agents could seek updated information about the emergency from the external source (i.e., become increasingly informed) as the simulation unfolded. Similarly, Yang et al., (2019) implemented communication via authorities sharing information to specific agents in the simulation, i.e., agents in certain households become informed about the emergency before others.

### 4.2.5. Use of communication categories over time

There are a very limited number of articles to assess patterns of prominent categories over time. However, models which implemented communicate between agents via social network links were most popular in 2022 (see Figure 5). A table of the prevalence of the categories between 2010 and 2022 is also provided in Appendix C.

Figure 5. The number of articles published using the communication categories between 2010 and 2022.

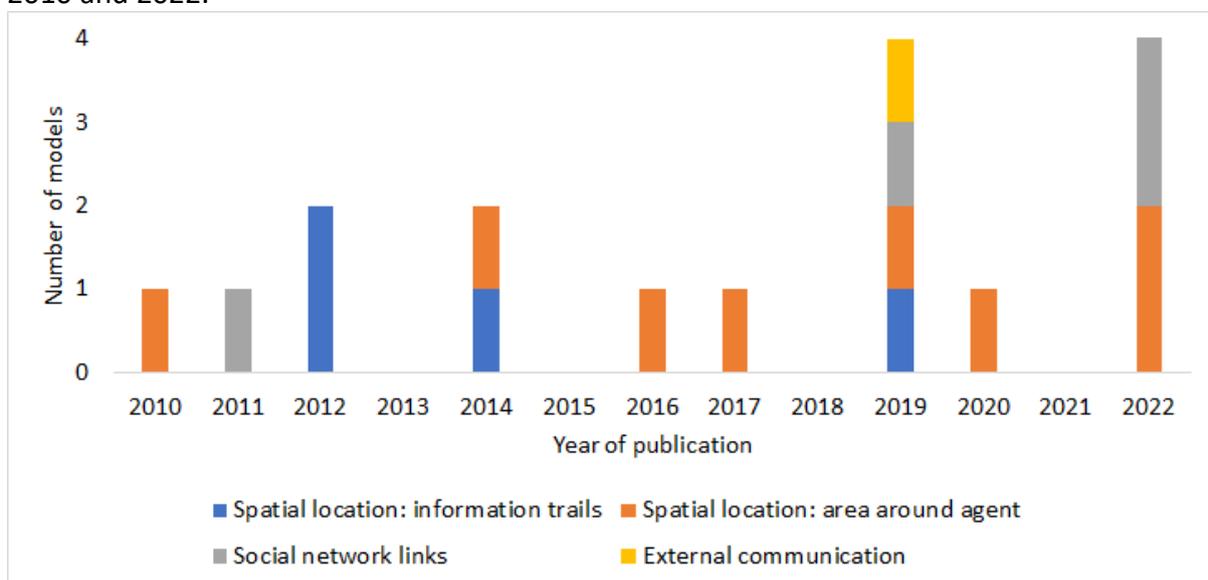

### 4.3. Common variables in agent-based evacuation models

We created four over-arching categories for the variables (see Table 6).

Table 6: Over-arching variable categories and their frequency in the models

| Over-arching category | Description | Number of articles | Percentage |
|---|---|---|---|
| Personal attributes | Individual agent attributes, e.g., demographic information, goals etc. | 62 | 88.6% |
| Information | Information exchanged between agents, e.g., quickest exit route etc. | 60 | 85.7% |
| Procedure | Processes agents incur during the emergency, e.g., distance travelled, path/route choice etc. | 46 | 65.7% |
| Environmental condition | Environmental conditions, e.g., congestion, door/exit width etc. | 11 | 15.7% |

We classified independent variables as those which may impact agent's decisions or performance during evacuation, e.g., it described the status and attributes of agents, behavioural features, influence of environment or other agents, and information available. Dependent variables are the outcome variables measured as a consequence of the independent variables, e.g., exit time or congestion in certain locations.

### 4.3.1. Personal attributes

Perhaps unsurprisingly given the focus on agent models, the most common variable category included in agent-based evacuation models was personal attributes. This included individual factors that were given to the agents (e.g., agent goals, see Guan et al., 2016) or were measured as an outcome of the simulation (e.g., agent emotions, see Li & Qin, 2012). The three most common variables were social influence, agent emotions, and agent roles. Table 7 provides a full list of the variables in the category.

Table 7: Independent and dependent variables in the personal attribute category.

| Personal attribute | Independent variable | Dependent variable | Combined | % |
|---|---|---|---|---|
| Social influence | 27 | 1 | 28 | 18.80% |
| Agent emotion | 19 | 5 | 24 | 16.10% |
| Agent roles | 18 | 1 | 19 | 12.80% |
| Demographics | 15 | 0 | 15 | 10.10% |
| Decision-making | 4 | 10 | 14 | 9.40% |
| Environment knowledge | 12 | 0 | 12 | 8.10% |
| Attraction/affiliation | 8 | 2 | 10 | 6.70% |
| Risk awareness | 10 | 0 | 10 | 6.70% |
| Travel speed | 3 | 7 | 10 | 6.70% |
| Agent goals | 7 | 0 | 7 | 4.70% |

### 4.3.2. Information

This category had one independent variable, which was where the source of the information given to the agents was manipulated. This was never used as a dependent variable. For example, Yang et al., (2019) provided information about a hurricane to agents in specific households, and then agents within the social network communicated who had decided whether to evacuate, informing the decisions of other agents to evacuate or remain.

### 4.3.3. Procedure

The second most popular variable category is procedure which refers to the process agents went through during the evacuation. For example, in some models the number of agents in the environment were manipulated (e.g., Guo et al., 2013), and in others the number of fatalities or number of exited agents were measured (e.g., Takabatake et al., 2017). The three most commonly used variables (in order of prevalence) were evacuation time, number of exited agents, and path or route choice of the agent. The full list of variables in this category are shown in Table 8.

Table 8: Independent and dependent variables in the procedure category.

| Procedure | Independent variable | Dependent variable | Combined | % |
|---|---|---|---|---|
| Evacuation time | 0 | 38 | 38 | 35.20% |
| Number of exited agents | 0 | 20 | 20 | 18.50% |
| Path/route choice | 1 | 16 | 17 | 15.70% |
| Exit choice | 2 | 14 | 16 | 14.80% |
| Number of agents | 4 | 5 | 9 | 8.30% |
| Fatalities | 0 | 6 | 6 | 5.60% |
| Distance travelled | 0 | 2 | 2 | 1.90% |

### 4.3.4. Environmental condition

The third most common variable category included manipulating or measuring the conditions in the environment. Environmental cues such as fire and smoke (e.g., Tissera et al., 2012) and density (e.g., Zhang et al., 2022) were the most manipulated and measured variables. Density was also a commonly measured outcome variable during the simulation (e.g., Bao & Huo, 2021), as was flow (e.g., Li & Han, 2015). The most frequently used variable in this category was density, followed by flow, and then congestion and environmental cues were equally the third most common. The list of variables in this category is shown in Table 9.

Table 9: Independent and dependent variables in the environmental category.

| Environmental condition | Independent variable | Dependent variable | Combined | % |
|---|---|---|---|---|
| Density | 11 | 9 | 20 | 25.30% |
| Flow | 6 | 12 | 18 | 22.80% |
| Congestion | 5 | 7 | 12 | 15.20% |
| Environmental cues | 12 | 0 | 12 | 15.20% |
| Distribution/placement | 6 | 1 | 7 | 8.90% |
| Door/exit width | 6 | 0 | 6 | 7.60% |
| Crowd pressure | 2 | 2 | 4 | 5.10% |

### 4.3.5. Prevalence of variables over time

We calculated the number of times variables within each overarching category were investigated in models between 2010 and 2022 (see Figure 6). This count combined both independent and dependent variables for each category. For example, in 2022, personal attributes were used 16 times as independent variables and 3 times as dependent variables, so the total number of times this category is included in 2022 is 19. However, the analysis for the independent variables and dependant variables separately are provided in Appendix D, as is as a table of the prominence of each category of variables used in the models across time.

Figure 6. The prominence of each category of variables used in the models between 2010 and 2022.

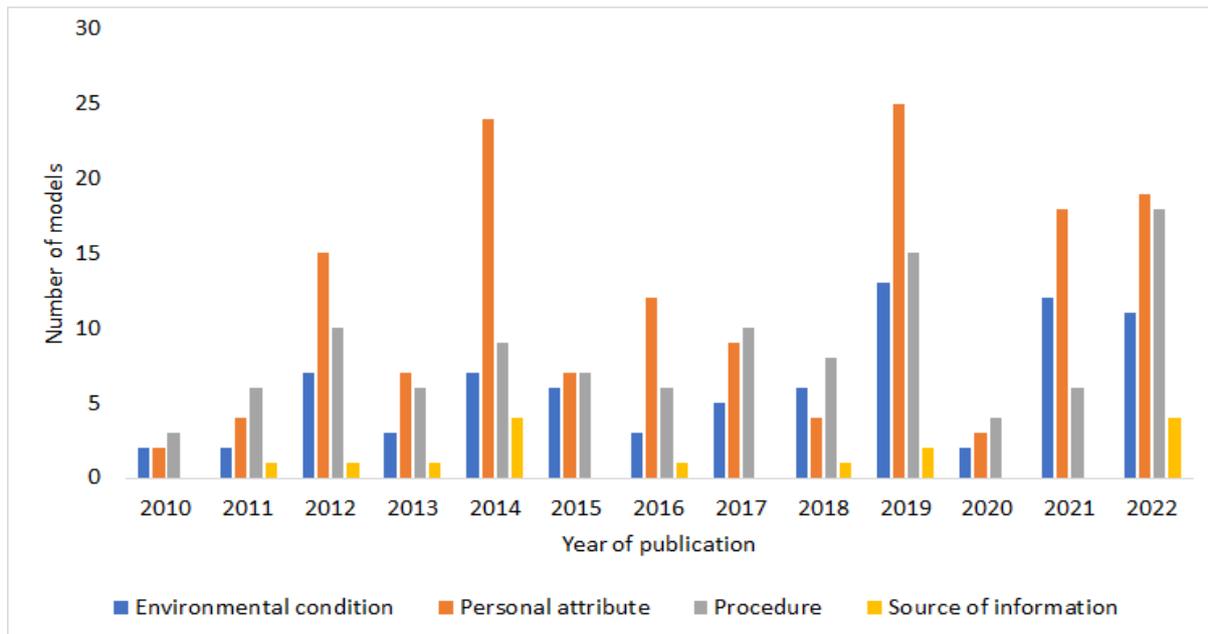

Since 2012, the consistently most prominent category is personal attributes. Within this category, social influence, agent emotions, and decision-making have become increasingly investigated since 2017 (see Figure 7).

Figure 7. Use of variables within the 'personal attributes' category between 2010 and 2022.

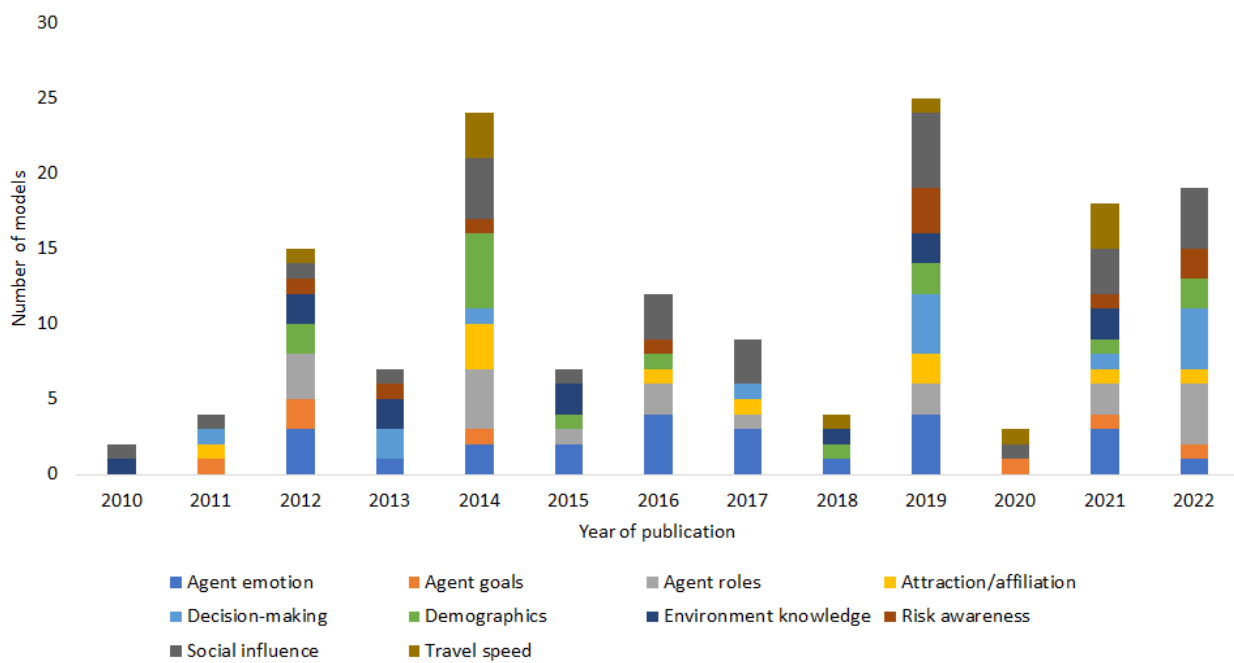

5    Discussion

This review identified how current agent-based evacuation models (published between 2010 and 2022) simulate social interactions and communication between agents, and the variables commonly used in these models. Agent-based evacuation models have diverse aims, purposes and areas of focus. However, we identified eight categories of social interactions among agents which ranged from focusing on physical components of interactions such as avoiding collisions, to socially connected agents that influenced one another's decision-making, emotions and behaviour. We also created four categories of how communication is transferred to and between agents during simulations: physically through either leaving "trails" of information in the environment or sharing information to nearby agents, through social networks, and receiving external information throughout the simulation. Finally, we identified the most common variables (components) used in the models and categorised these into overarching categories of personal attributes, procedures during the evacuation, environmental conditions, and information communication.

Here, we discuss the extent to which the categories of social behaviour and communication reflect empirical evidence of behaviour in emergencies. We also consider how the evidence suggests the most used variables in the evacuation literature may be affected by social behaviour or communication. Throughout these sections we recommend important evidence of human behaviours that predictive agent-based evacuation models can integrate to improve their realism, and the factors modellers should seek to investigate and integrate into their models to reflect case studies of emergencies.

5.1    Evidence and avenues for modelling social behaviour

Many of the agent-based evacuation models captured, at least to some extent, evidence of social interactions found in empirical research on human behaviour in emergencies. Models in the 'influence via information transfer' category aimed to simulate how information about the emergency was shared between people. This is a promising avenue for agent-based evacuation models since it reflects how people in real emergencies tend to seek and share information with others and decide response. For example, occupants of the World Trade Centre sought information from co-workers and supervisors when the towers were hit, and told others to evacuate (Averill et al., 2013). Similarly, members of the public who falsely believed an attack was occurring communicated with one another to establish the source of the threat (Drury et al., 2023).

Notably, the quality of the social bonds between people in emergencies influences whose information and actions are most influential. The basis of this idea is captured in the models allocated to the 'influence of pre-existing bonds' category where agents were influenced by the evacuation decisions of those in their networks. Specifically, the agents were most influenced by the decisions of those with whom they had the strongest relationships. Survivors of emergencies have reported coordinating with others because they felt part of the same social group (e.g., see Drury et al., 2009a), and feeling part of the same social group has been associated with collective behaviour such as expecting and providing support to others (Drury et al., 2016; Ntontis et al., 2018). People in ambiguous situations will particularly seek information from those with whom they already have prior positive relations (Templeton et al., 2023a) and the behaviour of fellow group members in ambiguous emergencies has been found to be particularly influential when deciding whether to flee from a potential threat (Drury et al., 2023).

However, only four articles based the social influence on pre-existing bonds between agents, making it the least common way of modelling social interactions. Going forward, modellers could consider instantiating a preference for information seeking from agents within the same group and reflect how their information and behaviour can be more influential than other unknown agents. These processes may affect outcomes commonly measured in the models. For example, evacuation time may be delayed because agents seek information, and their path or route choice may be affected by following others in their group, which may also impact density, flow and congestion if they move together.

The focus on social bonds in emergencies is an important route for agent-based evacuation models to simulate realistic social behaviour. However, some categories of social interactions in agent-based evacuation models are less evidence-based or require more sophisticated algorithms to reflect the empirical evidence. The social interaction categories 'influence via emotion transfer' and 'following' both instantiate an element of automatic influence between agents. The evidence from real world emergencies suggests that transference of influence is considerably more modulated in real emergencies.

In the 'influence via emotion transfer' category, a level of emotional transfer occurs between agents, primarily focusing on the spread of 'panic'. Although the effects of the 'panic' state on evacuation behaviour differs across models, the assumption that emotions automatically transfer to others (typically by close proximity to others) is not evidenced in the literature. For example, the notion of panic has been widely critiqued due to evidence that it is a very rare occurrence in emergencies (e.g., see Barr et al., 2022; Drury et al., 2013; Lorenz et al., 2017; Tierney et al., 2006). Based on the empirical evidence, we recommend that 'panic' behaviours or automatic influence be replaced by conceptualisations where panic is very rare, and influence is at least partially contingent on group membership.

For the 'following behaviour' category, agents begin to evacuate when they see others leaving. Recent research suggests that social influence – both for emotions and behaviour – is bounded by group relations. That is, people with whom a group bond is shared are more influential. For example, people were more likely to follow the same route others took in a maze when they believed that the others were in the same group as them (Neville et al., 2020). Similarly, residents of high-rise residential buildings were more likely to look to people in their group or who they had positive relations when deciding how to respond to a potential fire in their building (Templeton et al., 2023a). Thus, we recommend that when agent emotions are modelled, modellers should specify which emotions are being investigated and consider the role of group membership on how these emotions spread and develop over time. For modellers who seek to understand behaviour from previous emergencies, we recommend that they prioritise any available data on whose information the people in the emergency listen to and whose actions were followed during the emergency.

Finally, in the 'competition' category, agents competed for the same location and various methods are used to decide which agent received priority to reach the location. In routine scenarios, some individuals may compete for space, for example those in a hurry to overtake others. In those scenarios, competition variables can support a more realistic range of emergent outcomes. However, although there can be competition for space in emergencies, such as to avoid crowd crushes, these are typically due to external factors such as mistakes in planning and crowd management rather than intrinsic traits or decisions of the people in the emergencies (for a review, see Almeida & Schreeb, 2019).

In reality, cooperation and helping behaviour is highly common in emergencies where it is possible to provide it (for reviews, see Drury et al., 2020; for evidence see Ntontis et al., 2020; Drury et al., 2016). Importantly, group relations mitigate competition in emergency evacuations. Experimental research has suggested that people who most highly felt part of a group with others exhibited less competitive behaviour (e.g., pushing, shoving) compared to those with lower feelings of being part of the group (Drury et al., 2009b). Given that people tend to unite as a group when they believe they face the same emergency (e.g., Drury et al., 2009a), it follows that competition may be rare even when there were not previously bonds between people. We recommend that modellers evaluating previous case studies focus on how competition may be influenced by external factors (e.g., building layout, environmental factors) and seek to find and incorporate helping behaviour where appropriate. For predictive models, we recommend that helping behaviour is included as substantially more common than competitive behaviours in emergency scenarios, and to include the relationship between group membership and helping where possible.

## 5.2   Evidence and avenues for modelling communication

A particularly interesting finding in the models was the ways that communication between agents was implemented. Information about the emergency, environment, or actions of others was communicated either to nearby agents, through existing networks to agents that were not necessarily nearby, and externally through government, first responder, or news organisations. In many evacuations, people will attempt to gather information from multiple sources to establish what is happening and how to respond. This can include seeking nearby neighbours, communicating with others via phone and social media (e.g., Templeton et al., 2023a), and seeking information from the official news channels or professional response organisations (e.g., see Lindell et al., 2015).

The inclusion of multiple sources of information in models is a positive step (e.g., see Harris et al., 2022; Lovreglio et al., 2016). However, the views of the information provider may influence the extent to which people react to their information. Previous research into emergencies suggests that people are more likely to follow guidance from people and organisations that they feel are acting on their behalf (Frenzel et al., 2022), trust (Templeton et al., 2023b) and with whom they feel part of a group with (Carter et al., 2015b). However, these variables are quite dependent on broader contexts which might make them difficult to generalise for modelling scenarios. For example, seeing professional first responders as being part of the same group as the evacuees depends on what information was conveyed to the evacuees (e.g., what the emergency was, how to respond, and why, see Carter et al., 2015a). Trust in the information provider is also related to how much the evacuees believe the guidance is appropriate for their environment (Templeton et al., 2023b; Mayr et al., 2023).

More quantitative and qualitative research is needed in these areas to fully understand these dynamics and how they relate to evacuation behaviour, but for now predictive models could consider how adherence to evacuation guidance (e.g., path or route choice, number of exited agents) depends on who is providing the information, which information is given to the evacuees (e.g., why to take a specific route), and whether the guidance is sensible given the agent's knowledge of the environment.

In emergency evacuations, people often communicate about the source and severity of the threat, the environment (e.g., how to exit), and discuss how to respond. This was represented in the agent-based evacuation models we reviewed where agents shared information about risks in the environment (e.g., Zhang et al., 2014) and their knowledge of the environment (e.g., Henien & White, 2010). However, these models could be developed by considering how much the information from particular agents are trusted, and how this effects subsequent behaviour such as congestion or flow due to multiple agents communicating and taking the same route because they trust the information shared. Moreover, information is typically gathered from multiple sources and shared and discussed with others (Lindell et al., 2015). Going forward, predictive models could incorporate collective information gathering and decision-making on a larger scale, and consider how this impacts common variables such as overall evacuation time, social influence, and route or path choice.

Integrating these processes poses a key challenge for validation, particularly for how emotions such a trust affects behavioural outcomes. However, we suggest that future research seeks to test and validate these processes given the accumulating evidence showing their importance in guiding evacuation behaviour.

## 5.3. Strengths, missing literature, and limitations

This systematic review provides an overview of state-of-the-art agent-based evacuation models and synthesises the included literature to show common trends in how social interactions and communication are implemented, as well as the variables commonly explored across the models. By creating the systematic overview, we were able to evaluate how recent trends in the models compares to evidence on human behaviour in emergencies and suggest avenues forward to create more empirically based agent-based evacuation models.

However, there are limitations to this systematic review. As with any review, our search strategy, inclusion criteria and exclusion process may have failed to identity important articles. For example, we excluded interesting agent-based evacuation models that did not describe their underlying conceptual model – perhaps because the focus of the articles was to compare the capabilities of different software (e.g., Ren et al., 2022; Mu et al., 2014). Similarly, articles prior to 2010 were not included in order to capture more recent trends in the literature. For example, Proulx and Sime's (1991) seminal research on how communication approaches affect evacuation behaviour were not included due to the year of publication. Nevertheless, we argue that the knowledge and legacy of important articles prior to 2010 is still captured in the trends we identified in the current literature.

We are aware of prominent literature that was not included within our systematic review. There are articles which provide valuable insights into social interactions and communication but were not in evacuation contexts or that did not use agent-based models. One example is the seminal article by Moussaid et al. (2010) which uses observational data of walking behaviour in groups to create a social forces model based on social communication between group members. Another example is Kleinmeier et al. (2020) who created an agent-based model of cooperation in crowds, but was likely not identified during the search procedure due to not focusing on evacuations.

Similarly, although the included models used prominent pedestrian simulators such as PathFinder and MassMotion, none of the models used popular open-source pedestrian simulators such as the Jülich Pedestrian Simulator. Perhaps this is because the models were not focussed on evacuations, or the articles that did simulate evacuations were not published in the journals we used as filters in our research (e.g., Braun et al., 2019 published in EPiC Computing). Other popular simulators such as Vadere were also not included despite having publications that were highly relevant to the focus of the review, such as von Sivers et al. 2016 which focused on group processes and helping behaviour during evacuations. Possible reasons that the literature was not included are that we were restrictive when deciding our search terms and filtering criteria for subject areas and journals. However, we crafted these during the initial scoping phase of the search strategy to avoid excessive irrelevant articles. Future reviews could expand the search to include other types of models (e.g., social forces models), broader search terms, or opt for additional search methods such as a forward citation search or searching for connected paper.

Synthesising a broad field with varied aims and purposes mean that some articles fell into multiple categories. For example, Wang et al. (2015) fitted into two social interaction categories: competition and influence based on visual perception. Bulumulla et al., (2022) was allocated into two categories of communication: influence through spatial location and via social network. This is not necessarily a limitation but does reflect that the categories are not entirely independent and often models have multiple conceptual bases for agent behaviour.

A main challenge when creating the categories for social interactions and communication was that in many models – but not all models – the communication between agents influenced how other agents reacted. To make the separation between these categories clearer, we included articles in the 'influence via information transfer' social interaction category only if the information was shared by other agents in the environment and this directly affected agent behaviour. However, all models that integrated information sharing were included in the communication category, but the information was not always shared by fellow agents and did not always directly affect behaviour.

Another limitation is that it is difficult to decipher trends in the operationalisation of communication behaviour between 2010 and 2022 since there was a maximum of four articles published each year. A similar challenge was faced when identifying trends in the operationalisation of social interactions since only a handful of papers in each category were published per year.

Finally, the models included in the review included many variables such as the urban layout (Zlateski et al., 2020), behavioural uncertainty (Lovreglio et al., 2016), and group size (Barnes et al., 2021). Instead of including all variables, we identified the most commonly used variables in the literature to obtain prominent trends. However, we acknowledge that this does not provide an exhaustive list of all variables used in agent-based evacuation models. We recommend that any future reviews seek to address these limitations by testing the literature search criteria against the omitted articles described here, to ensure their inclusion in later iterations.

## 6  Conclusions

There are many promising recent developments in agent-based evacuation models that signal ways future models can integrate social interactions and communication between agents based on empirical evidence of behaviour in evacuations. The most promising models are those which model information communication between agents, between authority organisations and agents, and which instantiate influence between agents based on their strength of social connections. We argue that future agent-based evacuation models should attend to how information about the threat and environment are most trusted when coming from group members or those seen to be working on behalf of the group, and that the decisions and actions of fellow group members are most influential during evacuations. Further, we propose that future agent-based evacuation models should incorporate how these group processes impact common emergent outcomes such as evacuation time, number of exited agents, route and path choice, flow and congestion.

# Appendix A: Supplementary review of the commercial software

## 1. Introduction

Commercial crowd modelling software is frequently used to simulate evacuation scenarios, but these outputs are typically not published in academic literature. Nonetheless, their capabilities provide important information for what social interactions and communication can be captured in the models. Thus, we examined materials that described computer-based pedestrian tools that are either used or familiar to those in industrial practice.

Our objective was to review agent- or entity-based models, that are still actively promoted or supported, to determine the suitability of available functionality for the representation of communication between responder and pedestrians as examined elsewhere in this work[1].

There was a risk that if we only reviewed research publications in this supplementary task, we might miss the bulk of the models used in practice. This review was completed in a more ad hoc manner than the primary systematic review described in the main manuscript to capture the literature on the commercial software. The less systematic approach applied here was required given the frequent use of grey literature to describe commercial models, the use of web-based publishing to describe some of the models, and the absence of a definitive description of the frequency with which models are used.

## 2. Methodology

A snowball sampling approach was adopted starting from a baseline set of documents (see Primary References) that then identified models of particular interest or use. These include 1) existing model reviews which reflected pedestrian and evacuation models used in practice or that are proposed for such use, 2) grey literature reviews / reports, 3) practitioner surveys, 4) thesis/research reviews, and 5) public-facing websites where model descriptions are available.

A total of 83 models were identified from this review. Once these documents were reviewed, second-order references were identified that directly addressed model capabilities (refer to the Secondary References section below).  These were filtered according to whether 1) reference had been made to the model within the last 10 years, 2) the model was cited as still being in use (e.g., in a survey), 3) it was a pedestrian agent-based tool, and 4) the article described a functioning model, rather than some aspect of it or a platform for the development of it.

Models were then documented according to these attributes and other administrative elements, including model name, review source, associated link(s) and related articles. 26 models remained after this filtering process.  Those 26 models were reviewed in more detail to identify the main characteristics of current capabilities, existing trends and lessons that might be learned.

These included EXODUS, Legion, MassMotion, Pathfinder, OCEAN-HiDAC, STEPS, uCrowds, and the models produced by Durupınar et al., Xu et al., Wong et al., Jordao et al., Mao et al.,

He et al., Jiang et al., Gu et al., Qiu et al., Guy et al., Lemercier et al., Xu et al., Karamouzas et al., Jakin et al., Karamouzas et al., Jaklin et al., Mao et al., Bruenau et al., and Xu et al. (see Secondary References).

## 3. Results

The model capabilities are presented to demonstrate similarities between the findings of the commercial review and the primary systematic review. A simple scale is used to reflect the frequency of the results produced across the commercial models: '1-2' models (rare), 'Several' models (minority of models), 'Most' models (majority of models), and 'All' models. These simple categories were used given the variety of ways in which the models were documented and the potential for us mischaracterizing individual model capabilities and therefore the broader state of the modelling field.

3.1. Commercial model capabilities

3.1.1 Personal attributes

The results produced are broadly equivalent to those produced in the academic literature reviewed which modelled agent motivation and tasks.

3.1.1.1. Representation of agent motivation

Several models reflected internal agent motivational states that affect changes in available agent actions, such as switching exit and responding early. Some agent-based models examined the simulations of agents in detecting local conditions and adapting their behaviour to them, such as through smoke and congestion) (e.g., EXODUS, Pathfinder).

3.1.1.2. Representation of agent tasks

1-2 of the agent-based models represented the impact of external objectives on internal knowledge. This occurred either directly ('seeing' exit or smoke) or indirectly ('seeing' a sign that points to another sign or exit), (e.g., EXODUS, Pathfinder).

3.1.2. Environmental condition

All of the agent-based models examined could represent pre-evacuation times (e.g., EXODUS, Pathfinder, MassMotion). This is typically derived from a distribution, user specification or generated from exposure and/or external conditions (e.g., smoke). Often this is the main behavioural representation – representing a collection of activities through a single delay.

3.1.3. Maintaining the group structure

1-2 of the agent-based models examined allow for group behaviour through agents being assigned an attribute that then affiliated them with other agents with the same attribute. This might affect response and/or objectives in some way (e.g., EXODUS, Pathfinder,

MassMotion). This included the formation, reformation, and/or proximity maintenance of group members along with the alignment of member objectives.

### 3.1.4. Influence via information transfer

Like the social interaction category 'influence via information transfer' in the primary review, 1-2 of the agent-based models allowed for basic communication between agents, such as affecting pre-evacuation time, exit awareness, exit use, adoption of tasks, and motivation levels. These typically aligned to group objectives and were also crudely sensitive to the perception of individual roles (e.g., Pathfinder, EXODUS).

### 3.1.5. Physical environment

Most models emphasize performance at scale. Most commercial applications involve large-scale and complex structures (e.g., stadia, transportation hubs, etc.). However, the occasional hesitance of model developers to describe the assumptions on which agent performance is based (as opposed to tasks that can be performed), means it was challenging to determine the underlying model beyond establishing discrete functions and/or capabilities.

### 3.1.6. Representation of physical space

All of the agent-based models examined can represent physical space in some form (coarse, nodal, continuous, etc.). Some use multiple levels of representation (e.g., hybrid approaches) to capture navigational planning. All agent-based models represent individuals that are sensitive to their physical surroundings. Some represent other composite agents formed from individual agents (e.g., groups).

## 4. Discussion

### 4.1. Social interactions between agents

In general, the agent-based models examined had far greater sensitivity to the physical conditions than the social landscape. No commercial model currently has a coherent conceptual model of social interactions derived from research on human behaviour in evacuations. Typically, they include behaviours or sensitivity related to external conditions, or a conceptual model that is derived from research regarding a component of the evacuee decision-making process (e.g., OCEAN-HiDAC, ESCAPES). No model currently directly represents the normative or cultural context other than physically or through a composite of distinct behavioural settings.

### 4.2. Communication between agents

The provision of communication and sensitivity to information is significantly under-represented. Either it is not represented, is reflected in simplified binary states, or its perception and use is independent of an agent's attributes (e.g., experience, role, sensory abilities, etc.). There is little reference to an agent's experience (requiring an internal repository of information) and therefore perception is typically affected by *contemporary* abilities, impairments, or situations alone, rather than requiring assimilation or comparison with existing information. Typically, external information is represented as a condition for an action rather than a contributor to an ongoing process. The modelling focus is typically on information content and its immediate impact rather than the format, channel, source, source role or credibility. Information is treated primarily as 'a means to an end' that produces an action, rather than having a set of properties and being of interest in and of itself.

**Primary references**

**Secondary / Model references**

**Appendix B: Journals used as filters during the search in Science Direct**

In ScienceDirect, we filtered results by the most relevant common journals. The journals were: Ad Hoc Networks; Atmospheric Research; Automation in Construction Building and Environment; Chaos, Solitons & Fractals Computers, Environment and Urban Systems; Fire Safety Journal; Future Generation Computer Systems; IFAC-PapersOnLine; Information Sciences; Interacting with Computers; International Journal of Disaster Risk Reduction; International Journal of Industrial Ergonomics; Journal of Building Engineering; Journal of Theoretical Biology; Knowledge-Based Systems; Mathematical and Computer Modelling; Neurocomputing; Physica A: Statistical Mechanics and its Applications; Physics Letters A; Procedia Computer Science; Procedia Engineering; Procedia Manufacturing; Process Safety and Environmental Protection; Safety Science; Simulation Modelling Practice and Theory; Socio-Economic Planning Sciences; Transportation Research Part B: Methodological; Transportation Research Part C: Emerging Technologies; Transportation Research Part D: Transport and Environment; Transportation Research Procedia; Trends in Cognitive Sciences Tunnelling and Underground Space Technology.

**Appendix C: Prevalence of social interaction categories and communication categories between 2010 and 2022.**

**Implementation of social categories between 2010 and 2022.**

| Social interaction category | 2010 | 2011 | 2012 | 2013 | 2014 | 2015 | 2016 | 2017 | 2018 | 2019 | 2020 | 2021 | 2022 | SUM |
|---|---|---|---|---|---|---|---|---|---|---|---|---|---|---|
| Collision avoidance | 0 | 0 | 3 | 2 | 1 | 2 | 0 | 1 | 3 | 4 | 1 | 3 | 0 | 20 |
| Competition | 0 | 0 | 0 | 0 | 1 | 0 | 2 | 1 | 1 | 0 | 0 | 2 | 1 | 8 |
| Following | 0 | 1 | 2 | 0 | 2 | 1 | 2 | 2 | 0 | 2 | 1 | 1 | 2 | 16 |
| Maintain the group | 0 | 0 | 0 | 0 | 4 | 0 | 0 | 0 | 0 | 1 | 1 | 1 | 0 | 7 |
| Influence via visual perception | 1 | 1 | 0 | 1 | 1 | 2 | 0 | 2 | 1 | 3 | 1 | 3 | 1 | 17 |
| Influence via information transfer | 1 | 0 | 1 | 0 | 1 | 1 | 1 | 0 | 0 | 2 | 0 | 0 | 4 | 11 |
| Influence via emotion transfer | 0 | 0 | 1 | 0 | 1 | 0 | 0 | 1 | 0 | 0 | 0 | 2 | 1 | 6 |
| Influence via pre-existing bonds | 0 | 1 | 0 | 0 | 0 | 0 | 1 | 0 | 0 | 1 | 0 | 0 | 1 | 4 |

**Implementation of communication between 2010 and 2022.**

| Communication category | 2010 | 2011 | 2012 | 2013 | 2014 | 2015 | 2016 | 2017 | 2018 | 2019 | 2020 | 2021 | 2022 | SUM |
|---|---|---|---|---|---|---|---|---|---|---|---|---|---|---|
| Spatial location: information trails | 0 | 0 | 2 | 0 | 1 | 0 | 0 | 0 | 0 | 1 | 0 | 0 | 0 | 4 |
| Spatial location: area around agent | 1 | 0 | 0 | 0 | 1 | 0 | 1 | 1 | 0 | 1 | 1 | 0 | 2 | 8 |
| Social network links | 0 | 1 | 0 | 0 | 0 | 0 | 0 | 0 | 0 | 1 | 0 | 0 | 3 | 5 |
| External communication | 0 | 0 | 0 | 0 | 0 | 0 | 0 | 0 | 0 | 1 | 0 | 0 | 1 | 2 |

**Appendix D: Prevalence of variable categories between 2010 and 2022.**

| Variable categories | 2010 | 2011 | 2012 | 2013 | 2014 | 2015 | 2016 | 2017 | 2018 | 2019 | 2020 | 2021 | 2022 | SUM |
|---|---|---|---|---|---|---|---|---|---|---|---|---|---|---|
| Personal attribute | 1 | 3 | 4 | 3 | 8 | 4 | 4 | 5 | 3 | 11 | 2 | 6 | 8 | 62 |
| Environmental condition | 1 | 1 | 4 | 2 | 5 | 4 | 2 | 4 | 3 | 8 | 1 | 6 | 5 | 46 |
| information | 0 | 0 | 0 | 1 | 3 | 0 | 1 | 0 | 1 | 2 | 0 | 0 | 3 | 11 |
| Procedure | 1 | 3 | 5 | 3 | 8 | 4 | 2 | 5 | 4 | 9 | 3 | 4 | 9 | 60 |